\documentclass[3p,10pt,twocolumn]{elsarticle}

\usepackage{hyperref}
\usepackage{amsmath}
\usepackage{epstopdf,graphicx}
\usepackage{times}
\usepackage{rotating}

\journal{Journal of Physics Letters A}

\begin{document}

\begin{frontmatter}

\title{Quantum Speed Limit For Mixed States Using an Experimentally Realizable Metric}


\author{Debasis Mondal\corref{mycorrespondingauthor}\fnref{myfootnote1}}
\fntext[myfootnote1]{debamondal@hri.res.in}
\cortext[mycorrespondingauthor]{Corresponding author}

\author{Arun Kumar Pati\fnref{myfootnote}}
\fntext[myfootnote]{akpati@hri.res.in}
\address{Quantum Information and Computation Group,\\
Harish-Chandra Research Institute, Chhatnag Road, Jhunsi, Allahabad, India}

\begin{abstract}
Here, we introduce a new 
metric for non-degenerate density operator evolving along unitary orbit and show that this is experimentally 
  realizable operation dependent metric on the quantum state space. Using this metric, we obtain
  the geometric uncertainty relation that leads to a new quantum speed limit (QSL). We also obtain a Margolus-Levitin bound and an improved Chau bound for mixed states. We propose how to measure this new distance and speed limit in quantum interferometry. Finally, we also generalize the QSL for completely positive trace preserving evolutions. 
\end{abstract}

\begin{keyword}
\texttt{visibility}\sep\texttt{quantum speed limit}\sep\texttt{CPTP maps}\sep\texttt{quantum metrology}\sep\texttt{Margolus-Levitin bound}\sep\texttt{Mandelstam-Tamm bound}
\end{keyword}
\date{\today}
\end{frontmatter}



\section{Introduction}
In recent years, various attempts are being made in the laboratory to implement quantum gates, which are basic building blocks of a quantum computer. Performance of a quantum computer is determined by how fast one can apply these logic gates so as to drive
the initial state to a final state.
 Then, the natural question that arises is: can a quantum state evolve arbitrarily fast? It turns out that   
 quantum mechanics limits the evolution speed of any quantum
 system. In quantum information, study of these 
 limits has found several applications over the years. Some of these include, but not limited, to quantum metrology, quantum chemical dynamics, quantum control
 and quantum computation.

 Extensive amount of work has already been done on the subject ``minimum time required to
reach a target state" since the appearance of first
 major result by Mandelstam and Tamm \cite{Mandel:1945a1}. However, the notion of quantum speed or speed of transportation of quantum 
state was first introduced by Anandan-Aharonov using the Fubini-Study metric \cite{AA} and subsequently, the
same notion was defined in ref. \cite{pati} using the Riemannian metric \cite{provost}. It was found that the speed of a quantum state on the projective Hilbert 
space is proportional to the fluctuation in the Hamiltonian of the system. Using the concept of Fubini-Study metric on the projective Hilbert space, a geometric meaning is 
given to the probabilities of a two-state system \cite{patijoshi}. Furthermore, it was shown that the quantum speed is directly related to the super current in the Josephson 
junction \cite{anandanpati}. In the last two decades, there have been various attempts made in understanding the geometric aspects of quantum evolution for pure as well 
for mixed states \cite{Vaidman,EberlySingh,Fleming,BauerMello,wkwoott,
Bhattacharyya,LeubnerKiener,GisSabelli,
 UffHilge,Anandan,
Uhlmann,Uffink,Pfeifer,slcaves,PfFrohlich,akp,HoreshMann,
MargolusLevitin,Pati,
 Soderholm,GiovannettiPRA,GiovannettiEPLJOB,Andrecut, GrayVogt,LuoZhang,Batle,Borras2006,ZielinskiZych,Zander,Andrews,
Kupferman,LevitinToffoli,Yurtsever,FuLiLuo,hfchau,pkok,Brody,Frowis,Ashhab,
marcin,Auden,taddei,campo, manab,fungone,deffnerprl,Deffner,hyderi1,hyderi,fungtwo}. The quantum speed limit (QSL) 
for the driven \citep{Deffner} and the non-Markovian \citep{deffnerprl} quantum systems is introduced using the notion of Bures metric \cite{bures}. Very recently, QSL for physical processes was defined by Taddei {\it et al.} in Ref. \citep{taddei} using the Bures metric and in the case of open quantum system the same
is introduced by Campo {\it et al.} in Ref. \cite{campo} using the notion of relative purity \cite{Auden}. In an interesting twist, it has been shown that QSL
for multipartite system is bounded by the generalized geometric measure of entanglement \cite{manab}. 

 It is worthwhile to mention that very recently, an experiment was reported \cite{cimma}, which is the
only experiment performed, where only a consequence of the QSL had been
tested and any experimental test of the speed limit itself is still lacking. In this paper, we introduce a new operation dependent metric, which can be measured experimentally in the interference 
 of mixed states. We show that using this metric, it is possible to define a new lower limit
 for the evolution time of any system described by mixed state undergoing unitary evolution. We derive the QSL using the
  geometric uncertainty relation based on this new metric. We also obtain a Margolus-Levitin (ML) bound and an improved Chau bound for mixed states using our approach.
  We show that this bound for the evolution 
time of a quantum system is tighter than any other existing bounds for unitary evolutions. Most importantly, 
we propose an experiment to measure this new distance in the interference of mixed states. We argue that 
the visibility in quantum interference is a direct measure of distance for mixed quantum states. Finally, we generalize the speed limit for the case of completely positive 
trace preserving evolutions and get a new lower bound for the evolution 
time using this metric. 

The organization of the paper is as follows. In section II, we define the metric for the density operator along unitary path. Then, we use this metric to obtain new and tighter time bounds for unitary evolutions in section III, followed by examples in section IV. In section V, we show that bounds are experimentally measurable. Section VI is for generalization of the metric and the time bounds for completely positive trace preserving(CPTP) maps followed by an example. Then, we conclude in section VII. 


\section{ Metric along unitary orbit}
Let ${\cal H}$ denotes a finite-dimensional Hilbert space and ${\cal L}({\cal H})$ is the set of linear
operators on ${\cal H}$. A density operator $\rho$ is a Hermitian, positive and trace class operator
that satisfies $\rho \geq 0$ and ${\text{\rm Tr}} (\rho) =1$. Let $\rho$ be a non-degenerate density 
operator with spectral decomposition $\rho=\sum_{k}\lambda_{k}\vert k\rangle\langle k\vert$, where $\lambda_k$'s are the eigenvalues and $\{\vert k\rangle\}$'s are the eigenstates. We consider a system at time $t_{1}$ in a state $\rho_1$. It evolves under a unitary 
evolution and at time $t_2$, the state becomes $\rho_2= U(t_2, t_1)
\rho_1 U^{\dagger}(t_2, t_1)$. Any two density operators that are connected 
by a unitary transformation will give a unitary orbit. If $U(N)$ denotes the set of $N\times N$ unitary matrices on ${\cal H}^{N}$, then for a given density operator $\rho$, the unitary orbit is defined by $\rho'=\{U\rho U^{\dagger}: U\in U(N)\}$. The most important notion that
has resulted from the study of interference of mixed quantum states is the
concept of the relative phase between $\rho_1$ and $\rho_2$ and the notion of visibility in the interference pattern. The relative phase is defined by \cite{pati1}
\begin{eqnarray}\label{eq:1}
\Phi(t_2, t_1) = {\text{\rm Arg}} {\text{\rm Tr}}[\rho_1 U(t_2, t_1)]
\end{eqnarray}
and the visibility is defined by
\begin{eqnarray}\label{eq:1a}
V =\vert {\text{\rm Tr}}[\rho_1 U(t_2, t_1)]\vert.
\end{eqnarray}
Note that if $\rho_{1}=|\psi_{1}\rangle \langle 
\psi_{1}|$ is a pure state and 
$\vert \psi_{1}\rangle= | \psi(t_{1})\rangle\rightarrow | 
\psi_{2}\rangle=| \psi(t_{2})\rangle=U(t_{2},t_{1}) | \psi(t_{1})\rangle$, then
$|{\text{\rm Tr}}(\rho_1 U(t_2, t_1))|^2 = |\langle \psi(t_1)|\psi(t_2) \rangle |^2$,
 which is nothing but the fidelity between two pure states. The quantity ${\text{\rm Tr}}[\rho_1 U(t_2, t_1)]$ represents the 
probability amplitude between $\rho_1$ and $\rho_2$, which are unitarily connected.
Therefore, for the unitary orbit $|{\text{\rm Tr}}(\rho_1 U(t_2, t_1))|^2$ represents the transition probability between
$\rho_1$ and $\rho_2$. 

All the existing metrics on the quantum state space give rise to the distance between two states independent of the operation.
Here, we define a new distance between two unitarily connected states of a quantum system. This distance not only depends on the states but also 
depends on the operation under which the evolution occurs. Whether a state of a system will evolve to another state depends on the Hamiltonian which 
in turn fixes the unitary orbit. Let the mixed state traces out an open unitary curve $\Gamma: t\in [t_1, t_2] \rightarrow \rho(t)$
in the space of density operators with ``end points'' $\rho_1$ and $\rho_2$. If the unitary orbit connects the state $\rho_1$ at time $t_1$ to 
$\rho_2$ at time $t_2$, then the (pseudo-)distance between them is defined by
\begin{equation}\label{eq:2}
D_{U(t_{2},t_{1})}(\Gamma_{\rho_1},\Gamma_{\rho_2})^2:=
   4(1-  |{\text{\rm Tr}}[\rho_1 U(t_2, t_1)]|^2 ), 
   \end{equation} 
which also depends on the orbit, i.e., $U(t_{2},t_{1})$.  We will show that it is indeed a metric, i.e., it satisfies all the axioms to be a metric. 

We know that for any operator $A$ and a unitary operator $U$, $|\text{\rm{Tr}}(AU)|\leq\text{\rm{Tr}}|A|$ with equality for $U=V^{\dagger}$, where $A=|A|V$ is the
 polar decomposition of $A$ \cite{nielsen}. Considering $A=\rho=|\rho|$, we get $|{\text{\rm Tr}}[\rho_1 U(t_2, t_1)]|\leq 1$. This proves the non-negativity, or separation axiom. It can also be shown
that $D_{U}(\Gamma_{\rho_1},\Gamma_{\rho_2})=0$ if and only if there is no evolution along the unitary orbit, i.e., $\rho_{1}=\rho_2$ and $U=I$. If there is no evolution along the unitary orbit, then we have $U(t_2,t_1)=I$, i.e., trivial or global cyclic evolution, i.e., $\rho_2=U(t_2,t_1)\rho_1U^{\dagger}(t_2,t_1)=\rho_1$, which in turn implies $D_{U}(\Gamma_{\rho_1},\Gamma_{\rho_2})=0$. To see the 
converse, i.e., if $D_{U(t_2,t_1)}(\Gamma_{\rho_1},\Gamma_{\rho_2})=0$, then we 
have no evolution, consider the purification. We have $D_{U(t_2,t_1)}
(\Gamma_{\rho_1},\Gamma_{\rho_2})=4(1-\vert\langle\Psi_{AB}(t_1)\vert\Psi_{AB}
(t_2)\rangle\vert^2)$ where
 $\vert\Psi_{AB}(t_2)\rangle=U_A(t_2,t_1)\otimes I_B\vert\Psi_{AB}(t_1)\rangle$ such that $\text{Tr}_{B}(|\Psi_{AB}(t_{1})\rangle\langle\Psi_{AB}(t_{1}) |)=\rho_{1}$ and $\text{Tr}_{B}(|\Psi_{AB}(t_{2})\rangle\langle\Psi_{AB}(t_{2}) |)=\rho_{2}$. In the extended Hilbert space, $D_{U(t_2,t_1)}(\Gamma_{\rho_1},\Gamma_{\rho_2})=0$ implies $\vert\langle\Psi_{AB}(t_1)
\vert\Psi_{AB}(t_2)\rangle\vert^2=1$ and hence, $\Psi_{AB}(t_1)$ and $\Psi_{AB}(t_2)$ are same up to $U(1)$ phases. Therefore, in the extended Hilbert space, $D_{U(t_2,t_1)}(\Gamma_{\rho_1},\Gamma_{\rho_2})=0$ if and only if there is no evolution. But in the original Hilbert space there are non-trivial cyclic evolutions for which $D_{U(t_2,t_1)}(\Gamma_{\rho_1},\Gamma_{\rho_2})\neq 0$ in spite of the fact that $\rho_{1}=\rho_{2}$. To prove the symmetry axiom,
 we show that the quantity $|{\text{\rm Tr}}[\rho_1 U(t_2, t_1)]|$ is symmetric with respect to the initial and the final states. In particular, we have
\begin{equation}\label{eq:3a}
|{\text{\rm Tr}}[\rho_1 U(t_2, t_1)]|=|{\text{\rm Tr}}[\rho_2 U(t_1, t_2)]|=|{\text{\rm Tr}}[\rho_2 U(t_2, t_1)]|.
\end{equation}
 To see that the new distance satisfies the triangle inequality, consider its purificaton.
Let $\rho_{A}(t_1)$ and $\rho_{A}(t_2)$ are two unitarily connected mixed states of a quantum system $A$. If we consider the 
purification of $\rho_{A}(t_1)$, then we have $\rho_{A}(t_1)=\text{\rm{Tr}}_B[\vert\Psi_{AB}(t_1)\rangle\langle\Psi_{AB}(t_1)\vert]$,
 where $\vert\Psi_{AB}(t_1)\rangle=(\sqrt{\rho_{A}(t_1)}V_{A}\otimes V_{B})\vert\alpha\rangle\in \cal{H}_{A}\otimes\cal{H}_{B}$, $V_A$, $V_B$ are local unitary operators and
$\vert\alpha\rangle=\underset{i}{\sum}\vert i^A i^B\rangle$. The evolution of $\rho_{A}(t_1)$ under $U_A(t_2,t_1)$ is equivalent to
the evolution of the pure state $\vert\Psi_{AB}(t_1)\rangle$ under $U_A(t_2,t_1)\otimes I_B$ in the extended Hilbert space. Thus, in the extended Hilbert space,
 we have $\vert\Psi_{AB}(t_1)\rangle
\rightarrow\vert\Psi_{AB}(t_2)\rangle=U_A(t_2,t_1)\otimes I_B\vert\Psi_{AB}(t_1)\rangle$.
So, the transition amplitude between two states is given by
$\langle\Psi_{AB}(t_1)\vert\Psi_{AB}(t_2)\rangle =\text{\rm{Tr}}[\rho_{A}(t_1)U_A(t_2,t_1)]$. This simply says that the expectation value of a unitary operator $U_A(t_2,t_1)$ in a mixed state is equivalent to the 
inner product between two pure states in the enlarged Hilbert space. Since, in the extended Hilbert space the purified version of the metric satisfies
the triangle inequality, hence the triangle inequality holds also for the mixed states. Thus, $D_{U(t_2,t_1)}(\Gamma_{\rho_1},\Gamma_{\rho_2})$ is a distance in the extended Hilbert space and a pseudo-distance in the original Hilbert space. If $\rho_1 $ and $\rho_2$ are two pure states, which are unitarily connected then our new metric is the Fubini-Study metric \cite{pati,AA,karol} on the 
projective Hilbert space ${\text{\bf CP}}({\cal H})$. 



Now, imagine that two density operators differ from each other in time by an infinitesimal amount, i.e.,
$\rho(t_1) = \rho(t)=\sum_{k}\lambda_{k}|k\rangle\langle k|$ and  $\rho(t_2) = \rho(t+dt)=U(dt)\rho(t)U^{\dagger}(dt)$. Then, the infinitesimal distance between them is given by
\begin{equation}\label{eq:5}
dD_{U(dt)}^{2}(\Gamma_{\rho(t_{1})},\Gamma_{\rho(t_{2})}) = 4(1-  |{\text{\rm Tr}}[\rho(t) U(dt)]|^2 ).
\end{equation}
If we use the time independent Hamiltonian $H$ for the unitary operator, then keeping terms upto second order, the infinitesimal distance (we drop the subscript) becomes
\begin{eqnarray}\label{eq:6}
&dD^2=\frac{4}{\hbar^2}[{\text{\rm Tr}}(\rho(t) H^2) - [{\text{\rm Tr}}(\rho(t) H)]^2 ] dt^2\nonumber\\=&\frac{4}{\hbar^2}[\sum_{k}\lambda_{k}\langle k|H^2|k\rangle-(\sum_{k}\lambda_{k}\langle k|H|k\rangle)^2]dt^2\nonumber\\=&\frac{4}{\hbar^2}[\sum_{k}\lambda_{k}\langle \dot{k}|\dot{k}\rangle-(i\sum_{k}\lambda_{k}\langle k|\dot{k}\rangle)^2]dt^2,
\end{eqnarray}
where in the last line we used the fact that $i\hbar|\dot{k}\rangle=H|k\rangle$. Therefore, the total distance travelled during an evolution along the unitary orbit is given by
\begin{equation}\label{eq:6a}
D_{tot} = \frac{2}{\hbar} \int_{t_1}^{t_2} ~ (\Delta H)_\rho~ dt,
\end{equation}
where $(\Delta H)_\rho$ is the uncertainty in the Hamiltonian of the system in the state $\rho$ and is defined 
as $ (\Delta H)_\rho^2 = [{\text{\rm Tr}}(\rho(t) H^2) - [{\text{\rm Tr}}(\rho(t) H)]^2 ]$. Thus, it is necessary and sufficient to have non-zero $\Delta H$ for quantum system to evolve in time.

\section{ Quantum speed limit with new metric}
We consider a system $A$ with mixed state $\rho_A(0)$ at time $t=0$, which evolves to $\rho_{A}(t_2)=\rho_{A}(T)$ under a unitary operator $U_A(T$). We define the Bargmann angle in terms of the purifications of the states $\rho_A(0)$ and $\rho_A(T)$ in the extended Hilbert space 
${\cal H_A \otimes \cal H_B}$,
i.e., $\vert\langle\Psi_{AB}(0)\vert\Psi_{AB}(T)\rangle\vert=\cos \frac{s_0}{2}$, 
 where $\vert\Psi_{AB}(T)\rangle=U_A(T,0)\otimes I_B\vert\Psi_{AB}(0)\rangle$. It has already been shown in the previous section that $\vert \text{\rm{Tr}}[\rho_A(0) U_A(T)]\vert=\vert\langle\Psi_{AB}(0)\vert\Psi_{AB}(T)\rangle\vert$. Therefore, we can define the Bargmann angle between $\rho_{A}(0)$ and $\rho_{A}(T)$ as
\begin{eqnarray}\label{eq:6b}
\vert \text{\rm{Tr}}[\rho_A(0) U_A(T)]\vert=\cos \frac{s_0}{2},
\end{eqnarray}
such that $s_0\in [0, \pi]$. The quantity $s_{0}$ is nothing but the arccos of the visibility of the interference pattern between the initial and the final states and should not be confused with the Bures angle. It is shown later in this section that $s_{0}$ is never less than the Bures angle.

 We know that for pure states $\frac{2}{\hbar} \int ~ (\Delta H)_{|\psi_{AB}(0)\rangle}~ dt\geq \cos^{-1}|\langle\psi_{AB}(0)|\psi_{AB}(T)\rangle|$ as was derived in \cite{Mandel:1945a1,AA}. The inequality in the extended Hilbert space now becomes a property of the state space, i.e., $\frac{2}{\hbar} \int ~ (\Delta H)_\rho~ dt\geq s_0$. This says that the total distance travelled by the density operator $\rho(t)$ as measured by the metric (\ref{eq:6a}) is greater than or equal to the shortest distance between $\rho(0)$ and $\rho(T)$ defined by $s_{0}$.
 Using the inequality and the fact that the system Hamiltonian $H$ is time independent, we get the time limit of the evolution as
\begin{eqnarray}\label{eq:9}
T\geq\frac{\hbar}{\Delta H}\cos^{-1}|\text{\rm{Tr}}[\rho_{A}(0)U_{A}(T)]|.
\end{eqnarray}
This is one of the central result of our paper with the help of the new metric.
This same idea can be extended for the quantum system with time dependent Hamiltonian. The speed limit in this case is given by
\begin{eqnarray}\label{eq:9a}
 T\geq\frac{\hbar}{\overline{\Delta H}}\cos^{-1}|\text{\rm{Tr}}[\rho_{A}(0)U_{A}(T)]|,
\end{eqnarray}
where $\overline{\Delta H}=(\frac{1}{T}\int_{0}^{T} \Delta H dt)$ is the time averaged energy uncertainty of the quantum system.
This may be considered as generalization of the Anandan-Aharonov geometric uncertainty relation for the mixed states. This bound is better and tighter than
the bound given in \citep{GiovannettiPRA, Deffner} and reduces to the time limit given by Anandan and Aharonov \citep{AA} for pure
states. There can be some states called intelligent states and some optimal Hamiltonians for which the equality may hold. But in general, it is highly non-trivial to
find such intelligent states \cite{HoreshMann,Pati}.

To see that Eq. (\ref{eq:9}) indeed gives a tighter bound, consider the following. We suppose that a system in a mixed state $\rho_A$ evolves to $\rho'_A$ under $U_A(t)$. 
Let $S$ and $S'$ are the sets of purificatons of $\rho_A$ and $\rho'_A$ respectively. In \citep{GiovannettiPRA,Deffner,Uhlmann}, time bound was given in terms of Bures metric \cite{nielsen,karol,bures,jozsa}, i.e., $
  \min_{\vert\Psi_{AB}\rangle,\vert\Phi_{AB}\rangle} 2\cos^{-1}\vert \langle\Psi_{AB}\vert\Phi_{AB}\rangle\vert
$ \cite{nielsen}, such that $\vert\Psi_{AB}\rangle\in S$ and $\vert\Phi_{AB}\rangle\in S'$. But in Eq. (\ref{eq:9}), the time bound is tighter than that given in 
\citep{GiovannettiPRA,Deffner,Uhlmann} in the sense that here the bound is in terms of $s_0$, i.e.,$s_0$= $
2\cos^{-1}\vert \langle\Psi_{AB}\vert\Phi_{AB}\rangle\vert$, such that $\vert\Phi_{AB}\rangle=U_A\otimes 
I_B\vert\Psi_{AB}\rangle$ and hence, $s_0$ is always greater than or equal to the Bures angle \cite{nielsen,karol,bures,jozsa} defined as 2$\cos^{-1} 
[\text{\rm{Tr}}\sqrt{\rho^{\frac{1}{2}}_{A}\rho'_{A}\rho^{\frac{1}{2}}_{A}}]$. However, if $\rho$ is pure then the time bound given using our metric and the Bures metric are the same.

We have defined here the QSL based on an operation dependant metric, whereas the speed limits that exist in literature are operation independent. Our result  can be experimentally measurable whereas the existing results \citep{GiovannettiPRA,Deffner,Uhlmann} including the QSL in \citep{hyderi} cannot be measured directly. This is because we do not know yet how to measure the Bures metric and Uhlmann metric experimentally.

Furthermore, using our formalism, we can derive a ML time bound \cite{MargolusLevitin} for the mixed state. Let us consider the system $A$ with
 a mixed state $\rho(0)$ at time t=0. Let $\rho(0)=\sum_{k}\lambda_{k}\vert k\rangle\langle k\vert$ be the spectral decomposition of $\rho(0)$ and it evolves under a 
unitary operator $U(T)$ to a final state $\rho(T)$. In this case, we have
\begin{equation}\label{eq:9b}
\text{\rm{Tr}}[\rho(0) U(T)]=\sum_{n}p_{n}(\cos \frac{E_{n}T}{\hbar}-i\sin\frac{E_{n}T}{\hbar}),
\end{equation}
where we have used $\vert k\rangle$=$\sum_{n}c_{n}^{(k)}\vert\psi_{n}\rangle$, and $\vert\psi_{n}\rangle$'s are eigenstates  of the Hamiltonian $H$ with $H\vert\psi_{n}\rangle$=
$E_{n}\vert\psi_{n}\rangle$, and $p_n$=$\sum_{k}\lambda_{k}\vert c_{n}^{(k)}\vert^2$ with $\sum_{n}p_{n}=1$. Using the inequality $\cos x\geq 1-\frac{2}{\pi}(x+\sin x)$ for $x\geq 0$, i.e.,
 for positive semi-definite Hamiltonian, we get 
\begin{equation}\label{eq:9c}
\text{\rm{Re}}[\text{\rm{Tr}}[\rho(0) U(T)]]\geq [1-\frac{2}{\pi}(\frac{T\langle H 
\rangle}{\hbar}+\sum_{n}p_{n}\sin\frac{E_{n}T}{\hbar})].
\end{equation}
Then, from Eq. (\ref{eq:9c}), we have
\begin{equation}\label{eq:9d}
T\geq\frac{\pi\hbar}{2\langle H\rangle}[1-Re+\frac{2}{\pi}Im],
\end{equation}
where $Re$ and $Im$ are real and imaginary parts of $\text{\rm{Tr}}[\rho(0)U(T)]$ and they can be positive as well as negative. Note that when $Re$ and $Im$ are negative, this can give a tighter bound.
 This new time bound for mixed states evolving under unitary evolution with non-negative Hamiltonian reduces to $\frac{h}{4\langle H\rangle}$, i.e., the ML \cite{MargolusLevitin} bound in the case of evolution from one pure state to its orthogonal state. Therefore, the time limit of the evolution under unitary operation with Hamiltonian $H$ becomes

\begin{equation}\label{eq:9e}
T\geq\begin{cases}\max\{\frac{s_{0}\hbar }{2\Delta H},\frac{\pi\hbar}{2\langle H \rangle}(1+\frac{2}{\pi}Im-Re) \quad \phantom{0}       ~~~~~~\text{if}\, \text{ $H\geq 0$}\\
   \frac{\hbar s_0}{2\Delta H} \quad \phantom{0}~~~~~~~~~~~\text{otherwise.} 
          \end{cases}
    \end{equation} 
    
Similar results have also been shown in \cite{hyderi}. But these bounds can further be improved using an improved Chau \citep{fungone,hfchau} bound for mixed states using our formalism.
    Using the inequality $|Re(z)|\leq|z|$ and a trigonometric inequality $\cos x\geq 1-A|x|$, where $A\approx 0.725$ as found in \citep{hfchau}, we get $V=|\text{\rm{Tr}}(\rho U)|\geq |\sum_{n}p_{n}\cos(\frac{E_{n}T}{\hbar})|\geq 1-\frac{AT}{\hbar}\sum_{n}p_{n}|E_{n}|$. Therefore, the time bound is given by
    \begin{eqnarray}
    T\geq \frac{(1-V)\hbar}{A<E>},
    \end{eqnarray}
where $<E>$ is the average energy. It can be further modified to get a tighter bound as given by
\begin{eqnarray}
T\geq T_{c}\equiv\frac{\hbar}{A}\frac{(1-V)}{E_{DE}},
\end{eqnarray}
where $E_{DE}$ is the average absolute deviation from the median (AADM) of the energy as defined by Chau, i.e., $E_{DE}=\sum_{n}p_{n}|E_{n}-M|$ with M being the median of the $E_{n}$'s with the distribution $p_{n}$. The above bound is tighter time bound than that given in Eq. (\ref{eq:9}) depending on the distribution formed by the eigenvalues of $H$ for a sufficiently small visibility ($V$) \citep{hfchau}. Moreover, this new bound is always tighter than the Chau bound \citep{hfchau}. This is because of the fact that $V\leq \text{\rm{Tr}}\sqrt{\rho^{\frac{1}{2}}\rho'\rho^{\frac{1}{2}}}$. 

In the following, we have taken an example in the two dimensional state space and shown that the inequalities in Eq. (\ref{eq:9e}) are indeed satisfied by the quantum system.


\section{ Example of speed limit for unitary evolution}
We consider a general single qubit state $\rho(0)=\frac{1}{2}(I+\vec{r}.\vec{\sigma})$, such that $\vert r\vert^2$ $\leq 1$. Let it evolves under a general unitary operator $U$, i.e., $\rho(0)\rightarrow\rho(T)=U(T)\rho(0)U^{\dagger}(T)$, where
$U=e^{-i\frac{a}{\hbar}(\hat{n}.\vec{\sigma}+\alpha I)}$, $a=\omega.T$ and the Hamiltonian $H=\omega(\hat{n}.\vec{\sigma}+\alpha I)$ ($\vec{\sigma}$=
($\sigma_1,\sigma_2,\sigma_3$) are the Pauli matrices and $\hat{n}$ is a unit vector). This Hamiltonian $H$ becomes positive semi-definite for $\alpha\geq 1$. It is easy to show, that for $\alpha=1$,
\begin{eqnarray}
s_0=2\cos^{-1}\left(\begin{aligned}&[\cos^2 \frac{a}{\hbar}-(\hat{n}.\vec{r})\sin^2\frac{a}{\hbar}]^2+
\nonumber\\&\cos^2\frac{a}{\hbar}\sin^2\frac{a}{\hbar}(1+\hat{n}.\vec{r})^2\end{aligned}\right)^{\frac{1}{2}},
\end{eqnarray}
where $\Delta H$=$\omega\sqrt{1-(\hat{n}.\vec{r})^2}$, $Re=[\cos^2 \frac{a}{\hbar}-(\hat{n}.\vec{r})\sin^2\frac{a}{\hbar}]$, $Im=-\cos\frac{a}{\hbar}\sin\frac{a}{\hbar}(1+\hat{n}.\vec{r})$ and $\langle H\rangle=\omega(1+\vec{r}.\hat{n})$. Using the inequality (\ref{eq:9e}), for $\alpha=1=\hbar=\omega$ and $a=\pi/2$, we get that
the initial state evolves to $\rho(T)=\frac{1}{2}(I+\vec{r'}.\vec{\sigma})$, where $\vec{r'}=(2n_{1}(\hat{n}.\vec{r})-r_1,2n_{2}(\hat{n}.\vec{r})-r_2,2n_{3}(\hat{n}.\vec{r})-r_3)$ with evolution time bound given by  
\begin{equation}
T\geq\max\{\frac{\cos^{-1}(\hat{n}.\vec{r})}{\sqrt{1-(\hat{n}.\vec{r})^2}},\frac{\pi}{2}\}=\frac{\pi}{2}.
\end{equation}
 This shows that the inequality is indeed tight (saturated).  For simplicity, we consider a state with parameters $\hat{n}=(\frac{1}{\sqrt{2}},\frac{1}{\sqrt{3}},-\frac{1}{\sqrt{6}})$ and $\vec{r}=(0,0,\frac{1}{2})$ as an example. Then the state $\rho(0)$ under the unitary evolution $U(T)$ becomes $\rho(T)$=$\frac{1}{2}(I+\vec{r'}.\vec{\sigma})$, such that $\vec{r'}=(-\frac{4\sqrt{3}}{15},\frac{\sqrt{2}}{15},-\frac{1}{6})$. Therefore, the time bound given by Eq. (\ref{eq:9e}) is approximately max[1.09, 0.86], i.e., 1.09 in the units of $\hbar=\omega=1$. But a previous bound \citep{GiovannettiPRA,Deffner} would give 
approximately 0.31. This shows that our bound is indeed tighter to the earlier ones.

In the sequel, we discuss how the geometric uncertainty relation can be measured experimentally. This is the most important implication of our new approach.

\section{Experimental proposal To measure speed limit}
\begin{figure}
\includegraphics[scale=0.85]{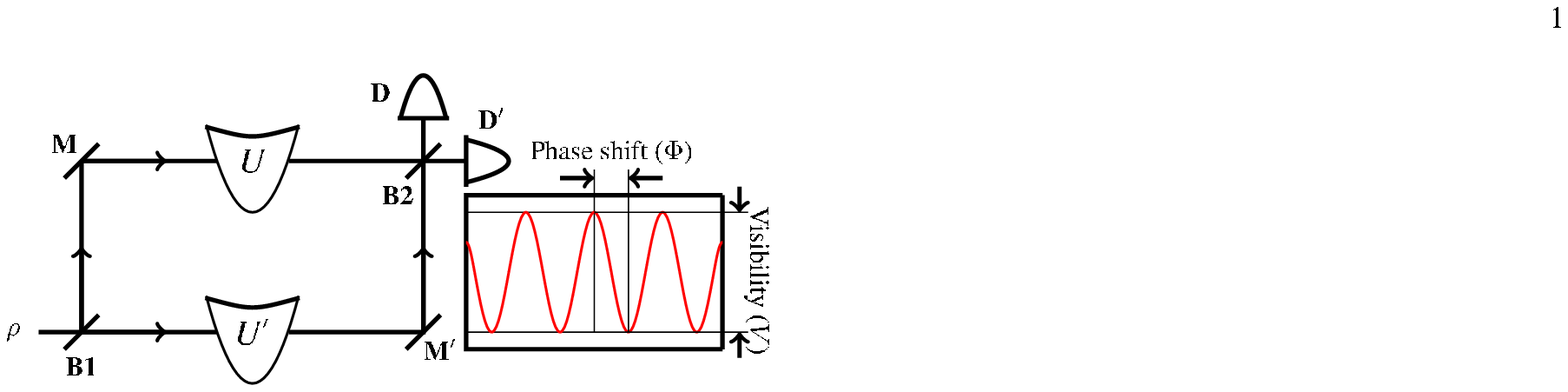}
\caption{Mach-Zender interferometer. An incident state $\rho$ is beamed on a 50\% beam splitter B1. The state in the upper arm is 
reflected through M and evolved by a unitary evolution $U$ and the state in the lower arm is evolved by an another unitary evolution $U'$ and then
 reflected through M$'$. Beams are combined on an another 50\% beam splitter B2 and received by two detectors D and D$'$ to measure the visibility. 
By appropriately choosing different unitaries, one can measure the quantum speed and the time limit.}
\label{fig:2}
\end{figure}

Arguably, the most important phenomenon that lies at the heart of quantum theory is the quantum interference. It has been shown that in the interference of mixed quantum states, 
the visibility is given by $V =\vert \text{\rm{Tr}}(\rho U)\vert$ and the relative phase shift is given by $\Phi =\text{Arg[\rm{Tr}}(\rho U)]$ \cite{pati1}. In quantum theory both of these play very important roles and they can be measured in experiments \cite{lett100403, lett050401}. The notion of interference of mixed states has been used to define interference of quantum channels \citep{daniel}. For pure quantal states, the magnitude of the visibility 
is the overlap of the states between the upper and lower arms of the interferometer. Therefore, for mixed states one can imagine that $\vert \text{\rm{Tr}}(\rho U)\vert^2$
also represents the overlap between two unitarily connected quantum states.
As defined in this paper, this visibility can be turned into a distance between $\rho$ and $\rho'=U \rho U^\dagger$. In Fig. \ref{fig:2}, we pass 
 a state $\rho$ of a system through a 50\% beam splitter B1.
  The state in the upper arm is reflected by M and evolved by a unitary evolution operator $U$ and 
the state in the lower arm is evolved by $U'$ and then reflected through M$'$. Both the beams in the upper and lower arms are combined on an another 
50\% beam splitter B2. The beams will interfere with each other. Two detectors are placed in the receiving ends and visibility of the interference
 pattern is measured by counting the particle numbers received at each ends. To measure the Bargmann angle, we apply $U=U(T)$ in one arm and $U'=I$ in 
another arm of the interferometer. The visibility $\vert\text{\rm{Tr}}[\rho(0)U(T)]\vert=\cos\frac{s_0}{2}$ will give the Bargmann angle $s_0$. To measure the quantum speed 
$v=\frac{2\Delta H}{\hbar}$, one can apply $U=  U(t)$ in one arm of the interferometer and one applies $U'=  U(t+\tau)$, where $\tau$ is very small in another arm 
of the interferometer. Then, the visibility will be $|{\text{\rm Tr}}[\rho(t)U(\tau)]|$. Hence, the quantum speed can be measured in terms of this 
visibility between two infinitesimally unitarily evolved states using the expression 
$V^2 = |{\text{\rm Tr}}[\rho(t)U(\tau)]|^2 = 1 - \frac{1}{4} v^2 \tau^2$. One can choose $\tau$ to be very short time scale with $\tau\ll T$. Once we 
know the visibility (the value of $s_0$) then we can verify the speed limit of the evolution for mixed states. Thus, by appropriately changing different unitaries, we 
can measure the quantum speed and hence the speed limit in quantum interferometry. One can also test the ML bound using our interferometric 
set up. Note that $R$ and $I$ of Eq. (\ref{eq:9e}) can be calculated from the relative phase $\Phi$ of quantum evolution together with the visibility $V$. The
 relative phase $\Phi$ of mixed state evolution can be measured by determining the shift in the interference pattern in the interferometer \cite{lett100403,lett050401}.
  Therefore, with prior knowledge of average of the Hamiltonian, one can test the ML bound for the mixed states. It should be noted that the measurement of the speed limit based on Bures metric requires the state tomography, whereas our bounds can be directly measured in the interference experiment and does not need to do the state tomography.

The notion of time bound can be generalized also for the completely positive trace preserving (CPTP) maps. In the next section, we derive the QSL for CPTP maps.

\section{ Speed limit under CPTP map}
 The metric defined in the paper gives the distance between two states which are
 related by unitary evolution. Now, consider a system $A$ in a state $\rho_A(0)$ at time $t=0$, which evolves under CPTP 
map ${\cal E}$ to  $\rho_{A}(T)$ at time $t=T$. The final state $\rho_{A}(T)$ can be expressed in the following Kraus operator representation form as 
\begin{equation}
\rho_{A}(T)={\cal E}(\rho_{A}(0))=\sum_{k}E_{k}(T)\rho_{A}(0)E^\dagger_{k}(T),
\end{equation} 
 where $E_{k}(T)$'s are the Kraus operators with $\sum_{k}E^\dagger_{k}(T) E_{k}(T)
=I$. We know that this CPTP evolution can always be represented as a unitary evolution in an extended Hilbert space via the 
Stinespring dilation. Let us consider, without loss of generality, an initial state $\rho_{AB}(0)=\rho_A
   (0)\otimes\vert\nu\rangle_B\langle\nu\vert$  at time $t=0$ in the extended Hilbert space. Here, $\nu$ is the initial state of the ancilla and it is in one of the basis state (of some chosen observable). The combined state evolves under 
   $U_{AB}(T)$ to a state $\rho_{AB}(T)$ such that $\rho_A(T)=\text{\rm{Tr}}_B[\rho_{AB}(T)]={\cal E}(\rho_{A}(0))$ and $E_{k}$=$_B\langle k\vert U_{AB}(T) 
\vert \nu\rangle_B$ \cite{karol}. Therefore, the time required to evolve the state $\rho_A(0)$ to $\rho_A(T)$ under the CPTP evolution is the same 
as the time required for the state $\rho_{AB}(0)$ to evolve to the state $\rho_{AB}(T)$ under the unitary evolution $U_{AB}(T)$ in the extended Hilbert space. One can also define the metric in the extended Hilbert space as
\begin{equation}
D^{2}_{U_{AB}}=4(1-|\text{\rm{Tr}}(\rho_{AB}(0)U_{AB}(T))|^2).
\end{equation}

Following the QSL for unitary case, we get the time bound to evolve the quantum system from $\rho_{A}(0)$ to $\rho_{A}(T)$ as
   \begin{equation}\label{eq:s11}
   T\geq\frac{\hbar s_0}{2\Delta H_{AB}},
   \end{equation}
   where $H_{AB}$ is the time independent Hamiltonian in the extended Hilbert space and $s_0$ is defined as
   \begin{equation}
     \cos\frac{s_0}{2}=\vert\text{\rm{Tr}}[\rho_{AB}(0)U_{AB}(T)]\vert.
   \end{equation}
 Note that the energy uncertainty of the combined system in the extended Hilbert space $\Delta H_{AB}$ can be expressed in terms of speed $v$ of evolution of the system and the Bargmann angle $s_0$ can be expressed in terms of operators acting on the  Hilbert space of quantum system. To achieve that, we express probability amplitude $ \text{\rm{Tr}}_{AB}[U_{AB}(T)(\rho_A(0)\otimes\vert\nu\rangle_B\langle\nu\vert)]$ in the extended Hilbert space in terms of linear operators acting on the Hilbert space of quantum system as
   \begin{eqnarray}\label{eq:s12}
& &\text{\rm{Tr}}_{AB}[U_{AB}(T)(\rho_A(0)\otimes\vert\nu\rangle_B\langle\nu\vert)]\nonumber\\&=&
     \text{\rm{Tr}}_A[\rho_{A}(0) E_{\nu}(T)],
\end{eqnarray}
where $E_{\nu}(T)$=$_B\langle\nu\vert U_{AB}(T)\vert\nu\rangle_B$. Here, $\vert\text{\rm{Tr}}_A[\rho_{A}(0) E_{\nu}(T)]\vert^2$ is the transition probability between the initial state and the final state of the quantum system under CPTP map. Note that the chosen $E_{\nu}$ depends on the initial ancilla state (even though the CP map does not depend on that). If the initial state of the ancilla is an arbitrary state $|e\rangle$, the Eq. \ref{eq:s12} can be expressed as $\text{\rm{Tr}}[U_{AB}(\rho_{A}(0)\otimes|e\rangle\langle e|)]=\text{\rm{Tr}}_{A}[E_{k}\rho_{A}(0)c_{k}^{*}]$, where $E_{k}=\langle k|U_{AB}|e\rangle$ and $c_{k}=\langle k|e\rangle$. Therefore, we can define the Bragmann angle between $\rho_{A}(0)$ and $\rho_{A}(T)$ under the CPTP map as 
\begin{equation}
\vert Tr_A[\rho_{A}(0) E_{\nu}(T)]\vert = \cos\frac{s_0}{2}.
\end{equation}
Similarly, we can define the infinitesimal distance between $\rho_{AB}(0)$ and $\rho_{AB}(dt)$ connected through unitary evolution $U_{AB}(dt)$ with time independent Hamiltonian $H_{AB}$ as
\begin{eqnarray}\label{eq:s13}
 dD^{2}_{U_{AB}(dt)}&=&4(1-\vert \text{\rm{Tr}}[\rho_{AB}(t) U_{AB}(dt)]\vert^2)\nonumber\\&=&4(1-\vert \text{\rm{Tr}}[\rho_{AB}(0) U_{AB}(dt)]\vert^2)\nonumber\\&=&4(1-\vert \text{\rm{Tr}}[\rho_{A}(0) E_{\nu}(dt)]\vert^2).
\end{eqnarray}
Now, keeping terms upto second order, we get the infinitesimal distance as
 \begin{equation}
dD^2 =\frac{4}{\hbar^2}[\text{\rm{Tr}}(\rho_{A}(0)\tilde{H^2}_{A})-[\text{\rm{Tr}}(\rho_{A}(0)\tilde{H}_{A})]^2] dt^2,
\end{equation}
where $\tilde{H}_{A}$=$_{B}\langle\nu\vert H_{AB}\vert\nu\rangle_{B}$ and $\tilde{H^2}_{A}$=$_{B}\langle\nu\vert H^{2}_{AB}\vert\nu\rangle_{B}$. Therefore, the speed of the quantum system is given by
\begin{equation}
v^2=\frac{4}{\hbar^2}[\text{\rm{Tr}}(\rho_{A}(0)\tilde{H^2}_{A})-[\text{\rm{Tr}}(\rho_{A}(0)\tilde{H}_{A})]^2].
\end{equation}
 Note, that this is not a fluctuation in $\tilde{H}_{A}$. This is because $\tilde{H_A}^2\neq\tilde{H_{A}^2}$. Here, $\tilde{H_{A}}$ can be regarded as an effective Hamiltonian for the subsystem $A$. Note that the speed can be expressed as 
\begin{eqnarray}
v^2=(\Delta \tilde{H})^2+\text{Tr}(\rho_{A}(0)\tilde{H^2})-\text{Tr}(\rho_{A}(0)\tilde{H}^2).
\end{eqnarray} 
 Hence, the time bound for the CPTP evolution from Eq. (\ref{eq:s11}) becomes 
\begin{equation}\label{eq:s14}
   T\geq\frac{2}{v}\cos^{-1}\vert \text{\rm{Tr}}_A[\rho_{A}(0) E_{\nu}(T)]\vert,
   \end{equation}
%

Here the interpretation of this limit is different from that of the unitary case. The transition probability in unitary
 case is symmetric with respect to the initial and the final states. Hence, the time limit can be regarded as the minimum time to evolve the initial state to the final state
 as well as the final state to the initial state. But the transition probability defined for positive map is not symmetric with respect to the initial and final states of 
the quantum system. In this case,
 time limit can only be regarded as the minimum time to evolve the initial state to the final state.
 
 Since we have mapped the time bound to evolve an initial state $\rho_{A}(0)$ to the final state $\rho_{A}(T)$ under CPTP evolution with the time bound of corresponding
 unitary representation $\rho_{AB}(T)=U_{AB}(T)\rho_{AB}(0)U^{\dagger}_{AB}(T)$ of the CPTP map in the extended Hilbert space, this speed limit can be measured in the 
interference experiment by interfering the two states $\rho_{AB}(0)$ and $\rho_{AB}(T)$ in the extended Hilbert space.
 
 We provide here an example of a general single qubit state $\rho_{A}(0)$=$\frac{1}{2}(I+\vec{r}.\vec{\sigma})$ at time $t$=0, such that $\vert r\vert^2$ $\leq 1$ evolving under CPTP map ${\cal E}$. It evolves to $\rho_{A}(T)$ at time $t$=$T$ under completely positive trace preserving (CPTP) map ${\cal E} : \rho_{A}(0)\rightarrow{\cal E}(\rho_{A}(0))$=$\rho_{A}(T)$=$\sum_{k}E_{k}(T)\rho_{A}(0)E^{\dagger}_{k}(T)$.  This evolution is equivalent to a unitary evolution of 
$ \rho_{AB}(0)=\frac{1}{2}(I+\vec{r}.\vec{\sigma})\otimes\vert 0\rangle\langle 0\vert\rightarrow\rho_{AB}(T)$ as
\begin{eqnarray}
 \rho_{AB}(T)=U_{AB}(T)\rho_{AB}(0)U^{\dagger}_{AB}(T)
\end{eqnarray} 
  in the extended Hilbert space. The unitary evolution is implemented by a Hamiltonian
 \begin{equation}
H=\sum_{i}\mu_{i}\sigma_{A}^{i}\otimes\sigma_{B}^{i}. 
\end{equation}  
 This is a canonical two qubit Hamiltonian up to local unitary operators. With the unitary $U_{AB}(T)$=$e^{-\frac{iT}{\hbar}(\sum_{i}\mu_{i}\sigma^{i}_{A}\otimes\sigma^{i}_{B})}$, we have the Kraus operators $E_{0}(T)$=$_B\langle 0\vert U_{AB}(T)\vert 0\rangle_{B}$ and $E_{1}(T)$=$_B\langle 1\vert U_{AB}
(T)\vert 0\rangle_{B}$ and it is now easy to show from Eq. (\ref{eq:s14}) that the time bound for this CPTP evolution is given by
\begin{eqnarray}\label{eq:s15}
T\geq\frac{\hbar \cos^{-1}K}{\sqrt{\mu^{2}_{1}+\mu^{2}_{2}+\mu^{2}_{3}(1-r^{2}_{3})-2\mu_{1}\mu_{2}r_{3}}},
\end{eqnarray}
where $K$=$[(\cos\theta_{1}\cos\theta_{2}\cos\theta_{3}+r_{3}\sin\theta_{1}\sin\theta_{2}\cos\theta_{3})^2+(\sin\theta_{1}\sin\theta_{2}sin
\theta_{3}+r_{3}\cos\theta_{1}\cos
\theta_{2}\sin\theta_{3})^2]^\frac{1}{2}$ and $\theta_{1}$=$\frac{\mu_{1}T}{\hbar}$, $\theta_{2}$=$\frac{\mu_{2}T}{\hbar}$ and $\theta_{3}$=$\frac{\mu_{3}T}{\hbar}$. If we consider $\theta_{1}$=$\pi$, $\theta_{3}$=$\pi$ then this bound reduces to $T\geq\frac{\hbar\theta_{2}}{\sqrt{\mu^{2}_{1}+\mu^{2}_{2}+\mu^{2}_{3}(1-r^{2}_{3})-2\mu_{1}\mu_{2}r_{3}}}$. One can also check our speed bound for various CPTP maps and it is indeed respected.

\section{Conclusion}
Quantum Interference plays a very important role in testing new ideas in quantum theory. Motivated by interferometric
set up for measuring the relative phase and the visibility for the pure state, we have proposed a new and novel measure of distance for the mixed states, which
are connected by the unitary orbit. The new metric reduces to the Fubini-Study metric for pure state. Using this metric,
we have derived a geometric uncertainty relation for mixed state, which sets a QSL for arbitrary unitary evolution. In addition, an ML bound and an improved Chau bound is derived using our formalism. These new speed limits based on our formalism are tighter than any other existing bounds. Since, the design of the target state is a daunting task in quantum control, our formalism will help in deciding which operation can evolve the initial state to the final state faster. Recently, an experiment was reported, which is the only experiment, where only a consequence of the QSL was tested and any experimental test of the QSL itself is still lacking. Here, we have proposed an experiment to measure this new distance and quantum speed in the interference of mixed states. The visibility of the quantum interference pattern is a direct measure of distance between two mixed states of the quantum system along the unitary orbit. We have shown that 
by appropriately choosing different unitaries in the upper and lower arm of the interferometer one can measure the quantum speed and the Bargmann angle. This 
provides us a new way to measure the quantum speed and quantum distance in quantum interferometry. We furthermore, extended the idea of speed limit
 for the case of density operators undergoing completely positive 
trace preserving maps. We hope that our proposed metric will lead to direct test of QSL in quantum interferometry. Our formalism can have implications in quantum metrology \cite{cable}, precision measurement of the gravitational red shift \cite{muller} and gravitationally induced decoherence \cite{zych} with mixed states and other areas of quantum information science.

{\bf Note:}  Recently, a number of papers have appeared on QSL relating it with the quantum coherence or asymmetry \cite{deba,marvian1}, leakage and decoherence \cite{marvian}, generation of non-classicality \cite{Jing} and also with the quantum Fisher information \cite{gerardo}. 

{\bf Acknowledgement:} DM acknowledges the research fellowship
of Department of Atomic Energy, Government of India


\begin{thebibliography}{70}
\bibitem{Mandel:1945a1}
L. Mandelstam and I. G. Tamm,
\newblock J. Phys. (Moscow) {\bf 9}, 249 (1945).
\bibitem{AA} J. Anandan and Y. Aharonov, Phys. Rev. Lett. {\bf 65}, 1697 (1990).
\bibitem{pati} A. K. Pati, Phys. Lett. A   {\bf 159}, 105  (1991).
\bibitem{provost} J. P. Provost and G. Valle, Commun. Math. Phys. {\bf 76}, 289 (1980).
\bibitem{patijoshi} A. K. Pati and A. Joshi, Euro. Phys. Lett. {\bf 21}, 723 (1993).
\bibitem{anandanpati} J. Anandan and A. K. Pati, Phys. Lett. A {\bf 231} 29 (1997).



\bibitem{Vaidman} L. Vaidman, Am. J. Phys {\bf 60}, 182 (1992).
\bibitem{EberlySingh} J. H. Eberly and L. P. S. Singh, Phys. Rev. D {\bf 7}, 359 (1973).
\bibitem{Fleming} G. N. Fleming, Nuovo Cimento A {\bf 16}, 263 (1973). 
\bibitem{BauerMello} M. Bauer and P. A. Mello, 
Annals of Physics {\bf 111}, 38 (1978).
\bibitem{wkwoott} W. K. Wootters, Phys. Rev. D {\bf 23}, 357 (1981).
\bibitem{Bhattacharyya} K. Bhattacharyya, J. Phys. A: Math. Gen. {\bf 16}, 2993 (1983).
\bibitem{LeubnerKiener} C. Leubner and C. Kiener, Phys. Rev. A {\bf 31}, 483 (1985).
\bibitem{GisSabelli} E. A. Gislason, N. H. Sabelli, and J. W. Wood, Phys. Rev. A {\bf 31}, 2078 (1985).
\bibitem{UffHilge}J. Uffink and J. Hilgevoord, Found. Phys. {\bf 15}, 925 (1985).
\bibitem{Anandan}J. Anandan, Found. Phys {\bf 21}, 1265 (1991).
\bibitem{Uhlmann} A. Uhlmann, Phys. Lett. A {\bf 161}, 329 (1992).
\bibitem{Uffink} J. Uffink, Am. J. Phys.  {\bf 61}, 935 (1993).
\bibitem{Pfeifer} P. Pfeifer, Phys. Rev. Lett. {\bf 70}, 3365 (1993).
\bibitem{slcaves} S. L. Braunstein and C. M. Caves, Phys. Rev. Lett {\bf 72}, 3439 (1994).
\bibitem{PfFrohlich} P. Pfeifer and J. Frolich, Rev. Mod. Phys. {\bf 67}, 759 (1995).
\bibitem{akp} A. K. Pati, Phys. Rev. A {\bf 52}, 2576 (1995). 
\bibitem{HoreshMann} N. Horesh and A. Mann, J. Phys. A: Math. Gen. {\bf 31}, L609 (1998).
\bibitem{MargolusLevitin} N. Margolus and L. B. Levitin, Physica D \textbf{120}, 188 (1998).
\bibitem{Pati} A. K. Pati, Phys. Lett. A {\bf 262}, 296 (1999).
\bibitem{Soderholm} J. Soderholm, G. Bjork, T. Tsegaye, and A. Trifonov, Phys. Rev. A {\bf 59}, 1788 (1999).
\bibitem{GiovannettiPRA} V. Giovannetti, S. Lloyd, and L. Maccone, Phys. Rev. A {\bf 67}, 052109 (2003).
\bibitem{GiovannettiEPLJOB} V. Giovannetti, S. Lloyd, and L. Maccone, Europhys. Lett. {\bf 62}, 615 (2003); J. Opt. B {\bf 6}, S807 (2004).
\bibitem{Andrecut} M. Andrecut and M. K. Ali, J. Phys. A: Math. Gen. {\bf 37}, L157 (2004).
\bibitem{GrayVogt} J. E. Gray and A. Vogt, J. Math. Phys. {\bf 46}, 052108 (2005).
\bibitem{LuoZhang} S. Luo and Z. Zhang, Lett. Math. Phys. {\bf 71}, 1 (2005).
\bibitem{Batle} J. Batle, M. Casas, A. Plastino, and A.R. Plastino, Phys. Rev. A {\bf 72}, 032337 (2005).
\bibitem{Borras2006} A. Borras, M. Casas, A. R. Plastino, and A. Plastino, Phys. Rev. A {\bf 74}, 022326 (2006).
\bibitem{ZielinskiZych} B. Zielinski and M. Zych, Phys. Rev. A {\bf 74}, 034301 (2006).
\bibitem{Zander}C. Zander, A. R. Plastino, A. Plastino, and M. Casas, J. Phys. A: Math. Theor. {\bf 40}, 2861 (2007). 
\bibitem{Andrews} M. Andrews, Phys. Rev. A {\bf 75}, 062112 (2007).
\bibitem{Kupferman} J. Kupferman and B. Reznik, Phys. Rev. A {\bf 78}, 042305 (2008).
\bibitem{LevitinToffoli} L. B. Levitin and T. Toffoli, Phys. Rev. Lett. {\bf 103}, 160502 (2009).
\bibitem{Yurtsever} U. Yurtsever, Phys. Scr. {\bf 82}, 035008 (2010).
\bibitem{FuLiLuo} S.-S. Fu, N. Li, and S. Luo, Commun. Theor. Phys. {\bf 54}, 661 (2010).

\bibitem{hfchau}H. F. Chau, Phys. Rev. A {\bf 81}, 062133 (2010).
\bibitem{pkok}P. J. Jones and P. Kok, Phys. Rev. A {\bf 82}, 022107 (2010).
\bibitem{Brody} D. C. Brody, J. Phys. A: Math. Theor. {\bf 44} 252002 (2011).
\bibitem{Frowis}F. Frowis, Phys. Rev. A {\bf 85}, 052127 (2012).
\bibitem{Ashhab} S. Ashhab, P. C. de Groot, and F. Nori, Phys. Rev. A {\bf 85}, 052327 (2012).
\bibitem{marcin}M. Zwierz, Phys. Rev. A {\bf 86}, 016101 (2012).
\bibitem{Auden} K. M. R. Audenaert, Quantum Info. Comput. {\bf 14}, 1-2 (January 2014), 31-38.
\bibitem{taddei}M. M. Taddei, B. M. Escher, L. Davidovich, and R. L. de Matos Filho, Phys. Rev. Lett. {\bf 110}, 050402 (2013).
\bibitem{campo} A. del Campo, I. L. Egusquiza, M. B. Plenio, and S. F. Huelga, Phys. Rev. Lett. {\bf 110}, 050403 (2013).
\bibitem{manab} M. N. Bera, R. Prabhu, A K Pati, A Sen(De), U Sen, arxiv:1303.0706.
\bibitem{fungone}Chi-Hang Fred Fung and H. F. Chau, Phys. Rev. A {\bf 88}, 012307 (2013).
\bibitem{deffnerprl} S. Deffner and E. Lutz, Phys. Rev. Lett {\bf 111}, 010402 (2013).
\bibitem{Deffner} S. Deffner and E. Lutz, J. Phys. A: Math. Theor. {\bf 46} 335302 (2013).
\bibitem{hyderi1}O. Andersson and H. Heydari, Entropy {\bf 15}, 3688 (2013).
\bibitem{hyderi}O. Andersson and H. Heydari, J. Phys. A:Math. Theor. {\bf 47}, 215301 (2014).

\bibitem{fungtwo}Chi-Hang Fred Fung and H. F. Chau, Phys. Rev. A {\bf 90}, 022333 (2014).
\bibitem{cimma}A. D. Cimmarusti, Z. Yan, B. D. Patterson, L. P. Corcos, L. A. Orozco, and S. Deffner, Phys. Rev. Lett. {\bf 114}, 233602 (2015).
\bibitem{pati1} E. Sjoqvist, A. K. Pati, A. Ekert, J. S. Anandan, M. Ericsson, D. K. L. Oi, and V. Vedral, Phys. Rev. Lett. {\bf 85}, 2845 (2000).\bibitem{nielsen} M. Nielsen and I. Chuang, $Quantum$ $Computation$ $and$ $Quantum$ $Information$, Cambridge University Press, 2000, 409-411.
\bibitem{karol} I. Bengtsson and K. Zyczkowski, $Geometry$ $Of$ $Quantum$ $States$, Cambridge University Press, 2006.
\bibitem{bures} D. Bures, Trans. Am. Math. Soc. {\bf 135}, 199 (1969).
\bibitem{jozsa} R. Jozsa, J. Mod. Opt. {\bf 41}, 2315 (1994).

\bibitem{lett100403} J. Du, P. Zou, M. Shi, L. C. Kwek, J. W. Pan, C. H. Oh, A. Ekert, D. K. L. Oi, and M. Ericsson, Phys. Rev. Lett. {\bf 91}, 100403 (2003).
\bibitem{lett050401} M. Ericsson, D. Achilles, J. T. Barreiro, D. Branning, N. A. Peters, and P. G. Kwiat, Phys. Rev. Lett. {\bf 94}, 050401 (2005).
\bibitem{daniel}D. K. L. Oi, Phys. Rev. Lett. {\bf 91}, 067902 (2003).


\bibitem{cable}K. Modi, H. Cable, M. Williamson and V. Vedral, Phys. Rev. X {\bf 1}, 021022 (2011).
\bibitem{muller}H. Müller, A. Peters and S. Chu, Nature {\bf 463}, 926–929 (2010).
\bibitem{zych}M. Zych, F. Costa, I. Pikovski and C. Brukner, Nat. Commun. 2:505 | DOI: 10.1038/ncomms1498 (2011).
\bibitem{deba}D. Mondal, C. Datta and S. Sazim, Phys. Lett. A {\bf 380} p. 689–695 (2016).
\bibitem{marvian1}I. Marvian, R. W. Spekkens and P. Zanardi, arXiv:1510.06474 (2015).
\bibitem{marvian}I. Marvian, D. A. Lidar, Phys. Rev. Lett. {\bf 115}, 210402 (2015).
\bibitem{Jing}J. Jing, L.-A. Wu, A. del Campo, arXiv:1510.01106 (2015).
\bibitem{gerardo}D. P. Pires, M. Cianciaruso, L. C. C\'eleri, G. Adesso and D. O. Soares-Pinto, arXiv:1507.05848 (2015).

\end{thebibliography}

\end{document}